\documentclass[12pt]{article}
\usepackage[latin1]{inputenc}
\usepackage{amssymb}
\usepackage{subfigure}
\usepackage{graphicx}
\usepackage{float}

\usepackage{amsmath}
\usepackage{color}

\newcommand{\bb}{\begin{equation}}
\newcommand{\ee}{\end{equation}}
\newcommand{\bega}{\begin{eqnarray}}
\newcommand{\ega}{\end{eqnarray}}
\newcommand{\begae}{\begin{eqnarray*}}
\newcommand{\egae}{\end{eqnarray*}}

\newcommand{\h}{\hspace*{4ex}}

\newcommand{\cent}{\centerline}
\newcommand{\vs}{\vspace*}

\begin{document}

\baselineskip 0.5cm

\begin{center}

{\large {\bf Optimizing optical trap stiffness for Rayleigh particles with a array of Airy beams } }

\end{center}

\vs{0.2 cm}

\cent{Rafael A. B. Suarez$^{\: 1}$, Antonio A. R. Neves$^{\: 1}$, and Marcos R. R. Gesualdi$^{\: 1}$}

\vs{0.2 cm}

\centerline{{\em $^{\: 1}$ Universidade Federal do ABC, Av. dos Estados 5001, CEP 09210-580, Santo Andr\'e, SP, Brazil.}}

\vs{0.5 cm}

{\bf Abstract  \ --} \ The Airy array beams are attractive for optical manipulation of particles owing to their non-diffraction and auto-focusing properties. An Airy array beams is composed of $N$  Airy beams which accelerate mutually and symmetrically in opposite direction, for different ballistics trajectories, that is, with different initial launch angles. Based on this, we investigate the optical force distribution acting on Rayleigh particles. Results show that is possible to obtain greater stability, for optical trapping, increasing the number of beams in the array. Also, the intensity focal point and gradient and scattering force of array on Rayleigh particles can be controlled through a launch angle parameter. \\


\vs{0.5 cm}

\h {\em\bf 1. Introduction} 

In 1986  A. Ashkin~\textit{et al}. were able to capture in three-dimension, dielectric particles using a single-beam tightly focused by a high numerical aperture lens. This technique is now referred to as ``optical tweezers'' or ``optical trapping''~\cite{Askin1986,Askin2000}. The measurement of forces on micron-sized particles of picoNewton order with high precision, optical tweezers has become a powerful tool for application in different fields of research, mainly in manipulation of biological system~\cite{Wang1997,Chiou2005}, colloidal systems, in nanotechnology for trapping of nano-structures~\cite{Marago2003}, in optical guiding and trapping of atoms~\cite{Kuga1997} as well as the study of mechanical properties of polymers and biopolymers.
\\
 
On the other hand, the study of non-diffracting waves or diffraction-resistant waves in optics regards special optical beams that keep their intensity spatial shape during propagation. Non-diffracting beams include Bessel beams, Airy beams, and others ~\cite{durnin1988comparison,Sivilo2007,Siviloglou2007,Suarez2016}; as well as the superposition of these waves can produce very special structured light beams, such as the Frozen Waves ~\cite{Vieira2012, Vieira2014}. These special optical beams present very interesting properties and could be applied in many fields in optics and photonics.
\\
 
Particularly the Airy beams (AiBs)~\cite{Sivilo2007,Siviloglou2007,Suarez2016} has attracted great interest recently in optical tweezers, for trapping and guiding of micro and nano-particles ~\cite{Baumgartl2008,Cheng2010,Yang2012,Zhao2016,Lu2017a}, due to their unusual features such as the ability to remain diffraction-free over long distances while they tend to freely accelerate during propagation~\cite{Sivilo2007,Siviloglou2007,Suarez2016}. An important parameter in the dynamic propagation is the initial launch angle~\cite{Siviloglou2008,Hu2010} which can be used to obtain optimal control of the ballistic trajectory. Recently, several authors have addressed the propagation properties of AiBs to study of circular Airy beams (CAiBs) ~\cite{Efremidis2010,Papazoglou2011,Jiang2018} and radial array Airy beams ~\cite{Vaveliuk2014,Chen2014} for optical trapping due to it unique abruptly auto-focusing characteristics where optical spot obtained in focal field can be used for simultaneous trapping of multiple particles~\cite{Zhang2011,Jiang2013,Jiang2016,Zhang2013,Cheng2014}.
\\

In this way, an Airy array beams (AiABs) are attractive for optical manipulation of particles owing to their non-diffraction and auto-focusing properties. An AiABs is composed of $N$ Airy beams which accelerate mutually and symmetrically in opposite direction, for different ballistics trajectories, that is, with different initial launch angles. This work presents the highlight that general AiABs can be optimized for maximum trap stiffness for specific beam parameters such as the initial launch angle ($\nu$) and the number of paired Airy beams in the array ($N$). The increase in launch angle and the number of beams increases the trap stiffness non-linearly.
 \\

\h {\em\bf 2. Theoretical background} 

\textbf{Airy beams (AiBs):} The solution for AiBs propagating with finite energy can be obtained by solving the normalized paraxial equation of diffraction in $1$D~\cite{Sivilo2007} 
\begin{equation}
i\frac{\partial}{\partial \xi}\psi\left(s,\xi \right)+\frac{1}{2}\frac{\partial^{2}}{\partial s^{2}}\psi\left(s,\xi \right)=0\,,
\label{paraxial_equation}
\end{equation}
where $\psi$ is the scalar complex amplitude, $s=x/x_{0}$ and  $\xi=z/kx_{0}^{2}$ are the dimensionless transverse and longitudinal coordinates, $x_{0}$ its characteristic length and $k=2\pi n/\lambda_{0}$ is the wave-number of an optical wave. The Eq.~(\ref{paraxial_equation}) admits a solution at $\xi=0$, given by~\cite{Siviloglou2008}
\begin{equation}
\psi\left(s,0 \right)=Ai\left(s \right)\text{exp}\left(as \right)\text{exp}\left(i\nu s \right)\,,
\label{Airy_finity}
\end{equation} 
where $\text{Ai}$ is the Airy function, $a$ is a positive quantity which ensures the convergence of Eq.~(\ref{Airy_finity}), thus limiting the infinity energy of the AiBs and $\nu$ is associated with the initial launch angle of this beam. The scalar field $\psi\left(s,\xi \right)$ is obtained from the Huygens--Fresnel integral, which is highly equivalent to Eq.~(\ref{paraxial_equation}) and determines the field at a distance $\xi$ as a function of the field at $\xi=0$~\cite{Siviloglou2008}, that is 
\begin{equation}
\begin{split}
\psi\left( s, \xi \right)&=Ai\left(s-\dfrac{\xi^{2}}{4}-\nu \xi + ia\xi\right)\text{exp}\left[ a \left(s - \dfrac{\xi^2}{2} - \nu \xi \right)\right] \\
	& \times  \text{exp}\left[ i \left( - \dfrac{\xi^3}{12} + \left(a^2 -\nu^2 + s \right)\dfrac{\xi}{2}  + \nu s -\nu \dfrac{\xi^2}{2} \right)\right].
\end{split}
\label{Airy_1D}
\end{equation}

This equation shows that the intensity profile decays exponentially as a result of modulating it with a spatial exponential function on the initial plane $\xi=0$. The term $s_{0}=s-\left( \xi^{2}/4\right)-\nu \xi$, where $s_{0}$ denotes the initial position of the peak at $\xi=0$, defines the transverse acceleration of the peak intensity of AiBs.

These results can be generalized for $2$D taking the scalar field of a beam described as the product of two independent components~\cite{Sivilo2007,Siviloglou2008}, that is:
\begin{equation}
\psi\left(s_{x},s_{y},\xi_{x},\xi_{y}\right)=\psi_{x}\left(s_{x},\xi_{x}\right)\psi_{y}\left(s_{x},\xi_{y}\right)\,,
\label{Airy_2D}
\end{equation}
where each of the components $\psi_{x}\left(s_{x},\xi_{x}\right)$ e $\psi_{y}\left(s_{x},\xi_{y}\right)$ satisfies the Eq.~(\ref{paraxial_equation}) and is given by Eq.~(\ref{Airy_1D}), with $s_{x}=x/x_{0}$, $s_{y}=y/y_{0}$, $\xi_{x}=z/kx_{0}^{2}$ e $\xi_{y}=z/ky_{0}^{2}$. To simplify
the AiBs description, we will consider a symmetrical configuration, such as, $a_{x}=a_{x}=a$, $\nu_{x}=\nu_{y}=\nu$ and $x_{0}=y_{0}=w_{0}$, resulting in $\xi_{x}=\xi_{y}=\xi=z/kw_{0}^{2}$.

\textbf{Airy array beams (AiABs):} Through a rotation of $\theta=2\pi/N$ in the transverse plane, we obtain a set of $N$ rotated AiBs. These evenly displaced AiBs on the transverse plane in $(\delta x,\delta y)$, accelerate mutually in the opposite direction~\cite{Cheng2014,Lu2017}. The AiABs can be described as
\begin{equation}
\Psi(s_{x},s_{y},\xi)=\sum_{j=1}^{N}\psi_{jx}(s_{jx},s_{jy},\xi)\psi_{jy}(s_{jx},s_{jy},\xi)\,, 
\label{matriz_superpositions}
\end{equation}
where the dimensionless transverse coordinates is given by
\begin{equation}
s_{jx}=\frac{(x\cos\theta_{j}-y\sin\theta_{j}+\delta x)}{w_{0}}\,, \quad s_{jy}=\frac{(x\sin\theta_{j}+y\cos\theta_{j}+\delta y)}{w_{0}}\,,
\label{matriz_superpositions1}
\end{equation}
The angle $\theta_{j}=2(j-1)\pi/N$ denotes the angle of rotation around the $z$ axis. 

\begin{figure}[H]
\includegraphics[scale=0.50]{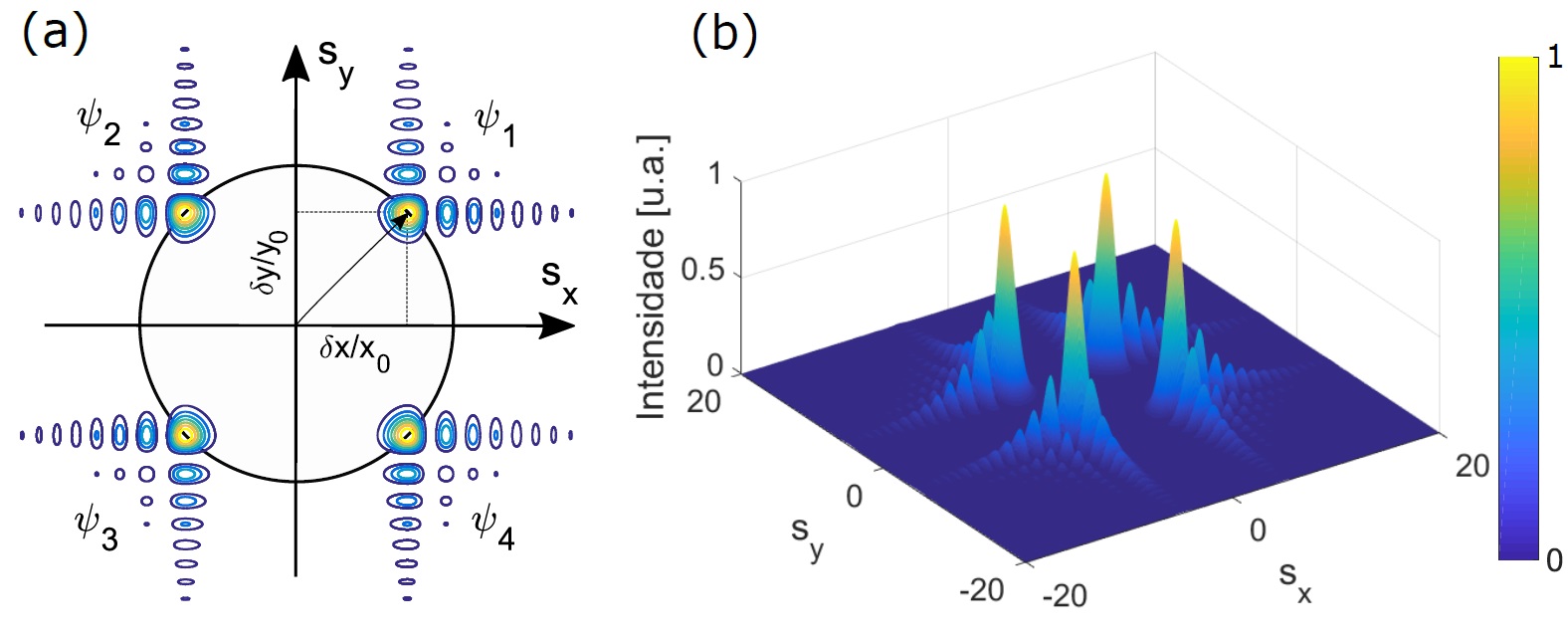}
\caption{$(a)$ Scheme for the generation of an AiABs with four AiBs ($N=4$). $(b)$ Normalized transverse intensity pattern from the plane $\xi=0$ onwards.}
\label{Rotation}
\end{figure}

The focal point $z_{f}=2kw_{0}(\sqrt{\nu^2+x_{0}/w_{0}} - \nu )$, is defined by the common point where the main lobes from all the beams in the array intercept along of the beam, where $x_{0}$ is the position of the first intensity peak. In the Fig.~\ref{Rotation} we can see a representative scheme of the generation process from four AiBs propagating along the $z$--axis.

\textbf{Electromagnetic fields:} We consider the following vector potential polarized in the $x$--direction
\begin{equation}
\textbf{A}=A_{0}\Psi e^{-ikz}\textbf{e}_{x}\,,
\label{potential_Airy}
\end{equation}
where $\Psi$ corresponds to the AiABs given by Eq.~(\ref{matriz_superpositions}) and $A_{0}$ is a normalization constant. Using the Lorenz gauge condition, the electromagnetic fields can be obtained from the vector potential $\textbf{A}$ as follows:
\begin{equation}
\begin{split}
\textbf{B}&=\nabla \times \textbf{A}\approx-ikA_{0}\left( \Psi \text{\textbf{e}}_{y} - \frac{i}{k}\frac{\partial\Psi}{\partial y}\text{\textbf{e}}_{z} \right) \text{e}^{-ikz}\,,\\
\textbf{E}=&-\frac{ic}{k}\nabla \times \textbf{B}\approx-ickA_{0} \left( \Psi \text{\textbf{e}}_{x} -\frac{i}{k}\frac{\partial\Psi}{\partial x}\text{\textbf{e}}_{z}\right) \text{e}^{-ikz}\,,
\end{split}
\label{EM_Field}
\end{equation}
where $c$ is the speed of light in vacuum, and which is later approximated in the paraxial limit.

Substituting Eq.~(\ref{matriz_superpositions}) into Eq.~(\ref{EM_Field}), the corresponding components of the magnetic and electric field are analytically expressed as
\begin{equation}
B_{y}=-ikA_{0}\text{e}^{-iF}\sum_{j=1}^{N} \text{Ai}\left(D_{jx} \right)\text{Ai}\left(D_{jy} \right)\text{e}^{\left( B_{jx}+iC_{jx}\right)}\text{e}^{i\left( B_{jy}+iC_{jy} \right)} \,,
\label{Magnetic_Field_Components_y}
\end{equation}
\begin{equation}
\begin{split}
B_{z}&=-\dfrac{A_{0}}{w_{0}} \text{e}^{-iF}\sum_{j=1}^{N} \left\lbrace \text{Ai}\left(D_{jx} \right) \left[\dfrac{\partial \text{Ai}\left(D_{jy} \right)}{\partial s_{jy}} +\alpha \cos \theta_{j} \ \text{Ai}\left(D_{jy}\right)\right] + \right.\\
&  \left.  \text{Ai}\left(D_{jy}\right) \left[\dfrac{\partial \text{Ai}\left(D_{jx} \right)}{\partial s_{jy}}-\alpha \sin \theta_{j}\ \text{Ai}\left(D_{jx}\right)\right]\right\rbrace \text{e}^{\left( B_{jx}+iC_{jx}\right)}\text{e}^{i\left( B_{jy}+iC_{jy} \right)}\,,
\end{split}
\label{Magnetic_Field_Components_z}
\end{equation}
\begin{equation}
E_{x}=-ickA_{0} \text{e}^{-iF}\sum_{j=1}^{N} \text{Ai}\left(D_{jx} \right)\text{Ai}\left(D_{jy} \right)\text{e}^{\left( B_{jx}+iC_{jx}\right)}\text{e}^{i\left( B_{jy}+iC_{jy} \right)}\,,
\label{Electric_Field_Components_x}
\end{equation}
\begin{equation}
\begin{split}
E_{z}&=-\dfrac{cA_{0}}{w_{0}} \text{e}^{-iF}\sum_{j=1}^{N} \left\lbrace \text{Ai}\left(D_{jx} \right) \left[\dfrac{\partial \text{Ai}\left(D_{jy} \right)}{\partial s_{jx}}+\alpha \sin \theta_{j}\ \text{Ai}\left(D_{jy}\right)\right] +  \right.\\
& \left.  \text{Ai}\left(D_{jy} \right) \left[\dfrac{\partial Ai\left(D_{jx} \right)}{\partial s_{jx}}+\alpha \cos \theta_{j} \ \text{Ai}\left(D_{jx}\right)\right]\right\rbrace \text{e}^{\left( B_{jx}+iC_{jx}\right)}\text{e}^{i\left( B_{jy}+iC_{jy} \right)}
\end{split}\,,
\label{Electric_Field_Components_z}
\end{equation}
where $B_{jm}=a\left(s_{jm}-\xi/2-\nu \xi\right)$, $D_{jm}=s_{jm}-\xi^2/4-\nu \xi + ia\xi$, $\alpha=a+i\left(\xi/2+\nu \right)$,~$C_{jm}=i\left[-\xi^3/12+\left(a^2-\nu^2+s_{jm}\right)\xi/2+\nu s_{jm} -\nu\xi^2/2  \right]$ e $F=k^{2}w_{0}^{2}\xi$.
\\

\textbf{Gradient and scattering forces:} When an object is much smaller than the wavelength of the incident light, i.e., $R<<\lambda$, the conditions for Rayleigh scattering are satisfied, and rigorous electromagnetic approach can be avoided~\cite{neves2019analytical}.  In this regime, the optical forces can be determined by considering the object as an induced electric dipole. The average force experienced in the presence of an external electromagnetic field is split into the optical gradient force $\textbf{F}_{\text{gd}}$, that is responsible for confinement in optical tweezers, and the optical scattering force, that it is due to transfer of momentum from the field to the particle as a result of the scattering and absorption process~\cite{Harada1996,Jones2015}
\begin{equation}
\langle\textbf{F}\rangle = \textbf{F}_{g}+\textbf{F}_{s}=\dfrac{1}{4}\text{Re}(\alpha)\nabla \vert \textbf{E} \vert ^2 + \frac{C_{pr}}{c}\textbf{S}\, ,
\label{force}
\end{equation}
where $\alpha=4\pi \epsilon_{2}R^3\, (\epsilon_{p}-\epsilon_{m})/(\epsilon_{p}+2\epsilon_{m})$ is a Clausius--Mossotti relation for a sphere of dielectric permittivity $\epsilon_{p}$ and radius $R$ in a medium with dielectric permittivity $\epsilon_{m}$~\cite{Jiang2016,Jones2015}, $\textbf{E}$ is the electric field, $\textbf{S}$ is the time-averaged Poynting vector and $C_{\text{pr}}$ is the radiation pressure cross-section of the particle. In the case of a small dielectric particle, $C_{\text{pr}}$ is equal to the scattering cross section $C_{\text{sc}}$ and is given by~\cite{Harada1996,Cheng2010}
\begin{equation}
C_{pr} =C_{sc}= \frac{8}{3}\pi k^4 R^6 \left(  \frac{m^2-1}{m^2+2} \right)^2 
\label{cross_section}
\end{equation}
where $m = n_{p}/n_{m}$ is the relative refractive index of the particle. 

The three components of the gradient and scattering force can be expressed as,
\begin{equation}
 \left( F_{g}\right)_{i} = \epsilon_{0}\pi n_{m}^2 R^3 \left( \frac{m^2-1}{m^2+2}\right) \partial_{i} \left[  \vert E_{x} \vert^2 + \vert E_{y} \vert^2 + \vert E_{z} \vert^2\right]
\label{gradient_force_x}
\end{equation}
\begin{equation}
 \left( F_{s}\right)_{i} = \left( \frac{n_{p}}{c}\right)C_{pr}S_{i}  
\label{Scattering_force_x}
\end{equation}
for $i=x,y,z$. \\

\h {\em\bf 3. Results and Discussion}  

For the simulations presented here, we consider an AiABs in two--dimensional that propagates along $z$--axis, where each individual beam in the array is characterize by the following typical parameters:  $\lambda=1064nm$, $a=0.05$, $w_{0}=2.5\lambda$, and $\delta x = \delta y=10w_{0}$. The input beam power of the AiABs in the initial plane $z=0$ is $P=1W$.

The validity of the paraxial approximation can be determined by monitoring the relative error between exact and paraxial results for the intensity distribution. The relative error is calculated on the $z$--axis because we are interested in studying the distribution of radiation forces along the beam axis, particularly on the focal point $z_{f}$. 
\begin{figure}[H]
 \centering
 \includegraphics[scale=0.50]{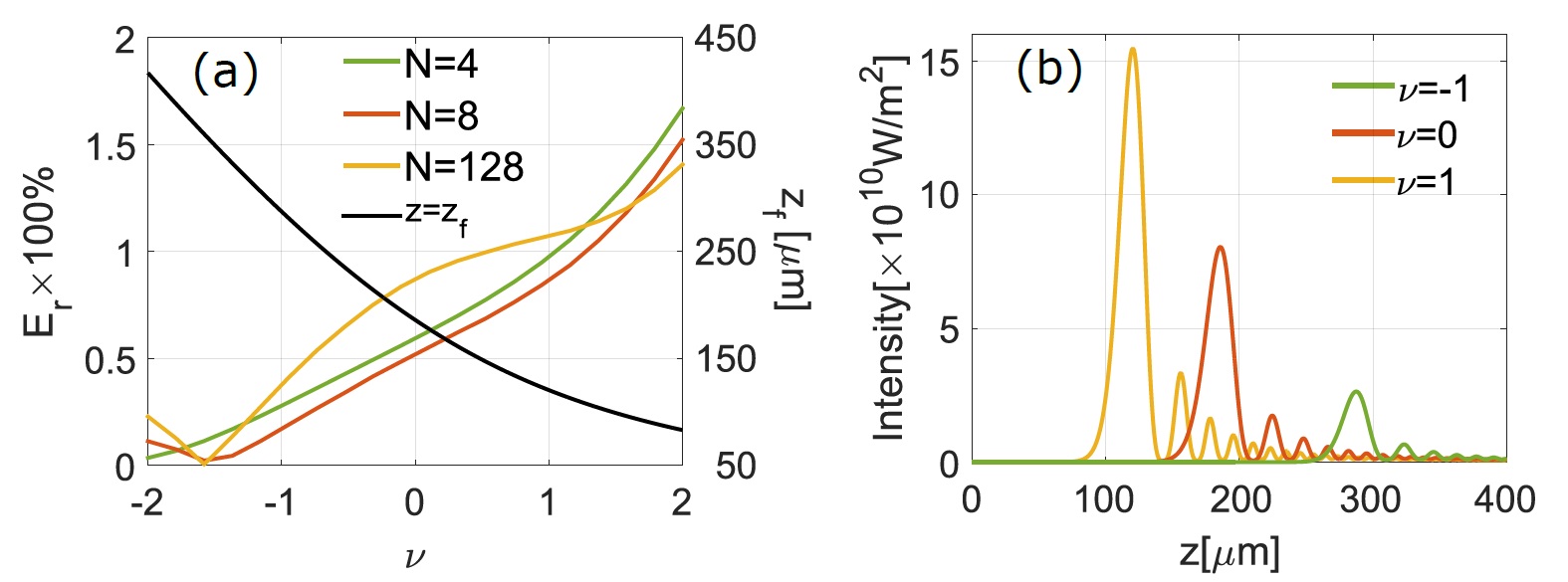}
 \caption{$\left(a\right)$ Relative error between exact and paraxial intensity distribution on the focal point as function of the initial launch angle $\nu$ for different $N$--values. The line black shows the focal point as function of $\nu$. $\left(b \right)$ Longitudinal intensity distribution.}
 \label{Error}
\end{figure}
In Fig.~\ref{Error}$(a)$, the relative error is determined as function of the $\nu$-parameter for different values of $N$, at the focal point $z_{f}$. We can see that the relative error increases with $\nu$ because for large values we move away from the paraxial condition. We can also observe that the focal point, change with $\nu$ and this is not affected by the number of beams in the array (black line). These results shows that we can use the paraxial approximation, with full validity, to describe the intensity distribution of the AiABs along the propagation axis for values of $\nu$ between $[-2,2]$.

In Fig.~\ref{Error}$(b)$, we shows the paraxial result of the longitudinal intensity for $N=128$ and different $\nu$--values. We can note that for negative values of $\nu$ the focal point increases and the intensity decreases, while for positive values the focal point decreases and the intensity increases. Therefore, we can have control of the intensity and the focal point simply by varying the initial launch angle.

\begin{figure}[H]
 \centering
 \includegraphics[scale=0.80]{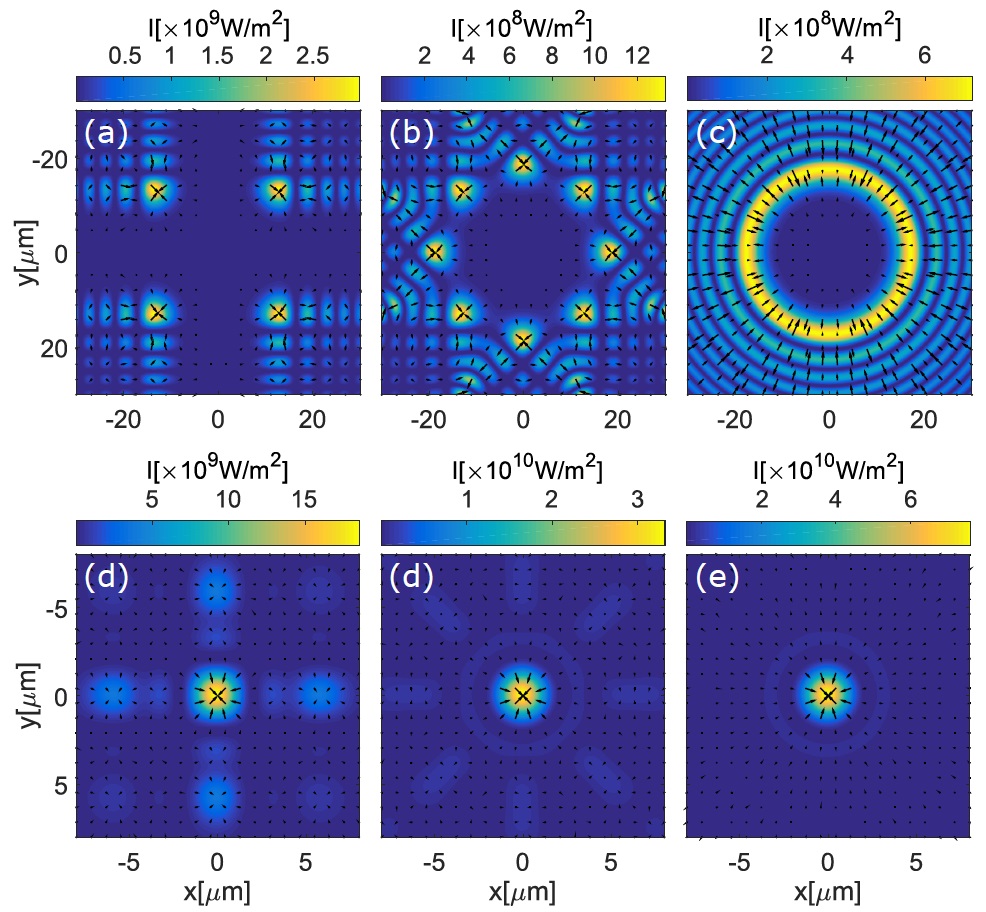}
 \caption{Direction and magnitude (arrows black) of transverse gradient force $(\textbf{F}_{\text{gd}})_\bot=(\textbf{F}_{\text{gd}})_{x}+(\textbf{F}_{\text{gd}})_{y}$ on planes $z=0$ (top row) an $z=z_{f}=185\mu m$ (bottom row) for: $N=4$ in $(a)$ and $(d)$, $N=8$ in $(b)$ and $(e)$, $N=128$ in $(c)$ and $(f)$. The cross-section intensity of the AiABs is also shown in the background of each frame.}
 \label{Force_distribution_different_N}
\end{figure}

\textit{Optical forces for different $N$-values:} We will examine the optical forces exerted on fused silica nano-particle of radius $R=50nm$ and refractive index $n_{p}=1.59$ in water $n_{m}=1.33$ when this is illuminated by an AiABs for different $N$--values and for $\nu=0$.

In Fig.~\ref{Force_distribution_different_N}, the direction and magnitude (arrows black) of the transverse gradient force $(\textbf{F}_{\text{gd}})_\bot=(\textbf{F}_{\text{gd}})_{x}+(\textbf{F}_{\text{gd}})_{y}$, is determined for the $z=0$ and $z_{f}=185.9\mu m$ plane. The cross-section intensity of the AiABs is also shown in the background of each frame. We can see that the focal intensity increases with the number of beams in the array (Fig.~\ref{Force_distribution_different_N}, left to right) to a maximum value $I_{m}=4.55\times 10^{10}W/m^2$ for $N=128$, always considering the same input power. This results implies that the particle can be transversely trapped in different intensity peaks on the plane $z=0$ and the transverse gradient force reaches its maximum value in $z_{f}$ because at this point the intensity is higher (e.g., for $N=128$, the maximum values of the focal intensity it's two orders of magnitude bigger than that in the $z=0$ plane).

\begin{figure}[H]
 \centering
 \includegraphics[scale=0.80]{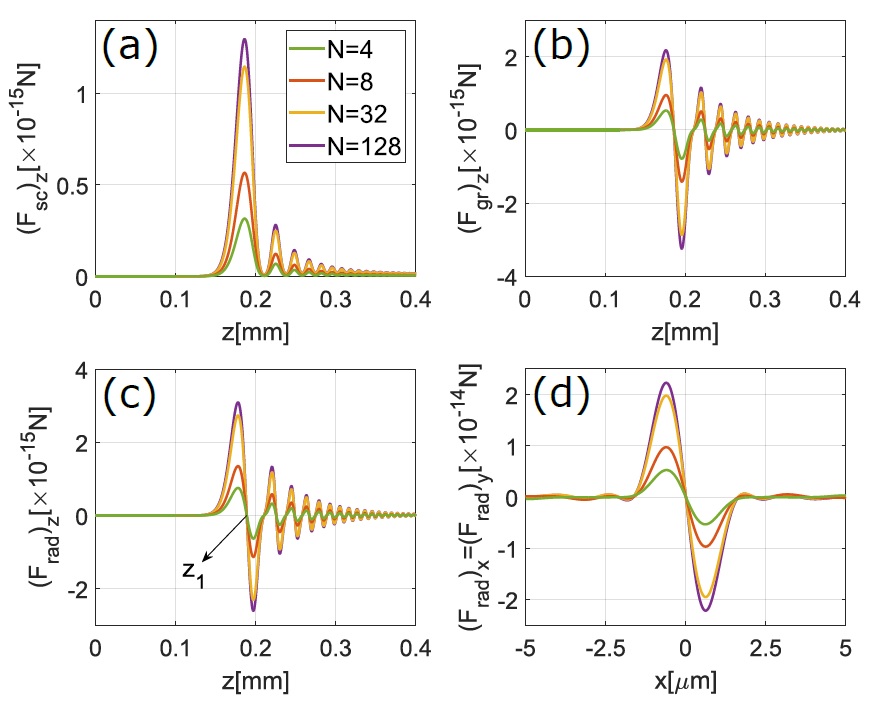}
 \caption{Optical forces along the $y=x$ direction for different values of $N$. $(a)$ Longitudinal scattering force. $(b)$ Longitudinal gradient force. $(c)$ Total radiation force $(\textbf{F}_{\text{rad}})_{z}=(\textbf{F}_{\text{gd}})_{z}+(\textbf{F}_{\text{sc}})_{z}$. $(d)$ The Transverse radiation force  $(\textbf{F}_{\text{rad}})_{x}=(\textbf{F}_{\text{gd}})_{x}+(\textbf{F}_{\text{sc}})_{x}$ in $z_{1}$.}
 \label{Intensity_Force}
\end{figure}

In Fig.~\ref{Intensity_Force}, the influence of different $N$ values in the array are determined for the force distribution along of the transverse line $y=x$. In Fig.~\ref{Intensity_Force} $(a)$ and $(b)$, the longitudinal component of the gradient and scattering forces are compared for different values of $N$ as a function of the on-axis position. The corresponding radiation force $(\textbf{F}_{\text{rad}})_{z}=(\textbf{F}_{\text{gd}})_{z}+(\textbf{F}_{\text{sc}})_{z}$ can be observed in Fig.~\ref{Intensity_Force} $(c)$. It shows that the particle can be trapped longitudinally at $z_{1}=188.1\mu m$ first point of stable equilibrium in the distribution of the longitudinal force. Note that the position of $z_{1}$ is slightly shifted from the focal point $z_f=185.9\mu m$ because of the influence of the scattering force.  The transverse component of the radiation force as a function of the transverse position in the planes $z_{1}$ is shown in Fig.~\ref{Intensity_Force} $(d)$. We can observe that the particle can be transversely trapped in these planes. However, the magnitude of the transverse radiation force gradually increases with $N$ until it reaches its maximum value $F_{x}=2.22\times 10^{-14}\text{N}$ for $z_{1}$.

\textit{Optical forces for different launch angle:} To determine the dependence on the optical forces due to different initial launch angles, $\nu-$~parameter, we employ an AiABs with $N=128$.

\begin{figure}[H]
 \centering
 \includegraphics[scale=0.80]{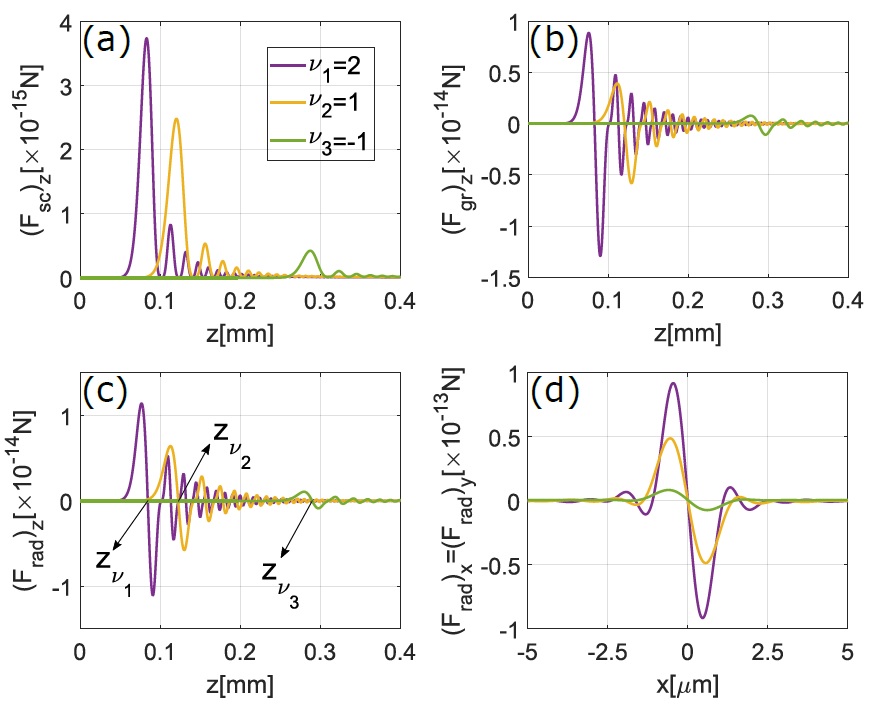}
 \caption{Longitudinal force distribution along $y=x$ for three different initial angle. $(a)$ Scattering force. $(b)$ Gradient force. $(c)$ Total radiation force. The points $z_{\nu=2}$, $z_{\nu=1}$ and $z_{\nu=-1}$ corresponds to the first stable equilibrium position along the longitudinal force profile. $(d)$ transverse radiation force in the first stable equilibrium position.}
 \label{Radiation_Forcas_nu_Diferent_z}
\end{figure}

It can be observed in Figs.~\ref{Radiation_Forcas_nu_Diferent_z} $(a)$ and $(b)$ the longitudinal component of scattering and gradient forces for three different values of launch angle parameter: $\nu_{1}=2$, $\nu_{2}=1$ and $\nu_{3}=-1$. This results revels that we can increasing the maximum intensity in the distribution and control the focal intensity point from the appropriate choose of the launch angle parameter $\nu$. The radiation force can be observed in Fig.~ \ref{Radiation_Forcas_nu_Diferent_z} $(c)$ where the first stable equilibrium points in the distribution occurs at the axial position $z_{\nu_1}=84.4\mu m$, $z_{\nu_2}=122.5\mu m$ and $z_{\nu_3}=289.2\mu m$ for $\nu_1$, $\nu_2$ and $\nu_3$, respectively. The Figs.~ \ref{Radiation_Forcas_nu_Diferent_z} $(d)$ shows the transverse radiation force as a function of particle position for three different values $\nu$ in the first stable equilibrium point respectively. As one can clearly see, that radiation force could by increases by choosing a positive launch angle. by contrast, this force can be reduced by taking negative values of $\nu$.

\emph{Trap stiffness} In first approximation, the radiation force may be expanded by a Taylor series around the equilibrium. This means that when the object is displaced from its equilibrium position, it experiences an attractive force that causes the particle to return to its equilibrium position. Therefore, along each direction, optical forces can be described by Hooke's law, where $\kappa_{x}=\partial_{x} F_{x}\vert_{x_{\text{eq}}}$, $\kappa_{y}=\partial_{y} F_{y}\vert_{y_{\text{eq}}}$, and $\kappa_{z}=\partial_{z} F_{z}\vert_{z_{\text{eq}}}$ are the trap stiffens along the $x-$, $y-$, and $z-$~directions, respectively.

The potential energy in the traps can be approximated by harmonic potential wells, that is 

\begin{equation}
 U=\dfrac{1}{2}\kappa_{\perp}r^2+\dfrac{1}{2}\kappa_{z}z^2
\label{stable_trapping}
\end{equation}
where $\kappa_{\perp}=\kappa_{x}+\kappa_{y}$ and $\kappa_{z}$ is the transverse and axial trap stiffness respectively. 

\begin{figure}[H]
 \centering
 \includegraphics[scale=0.80]{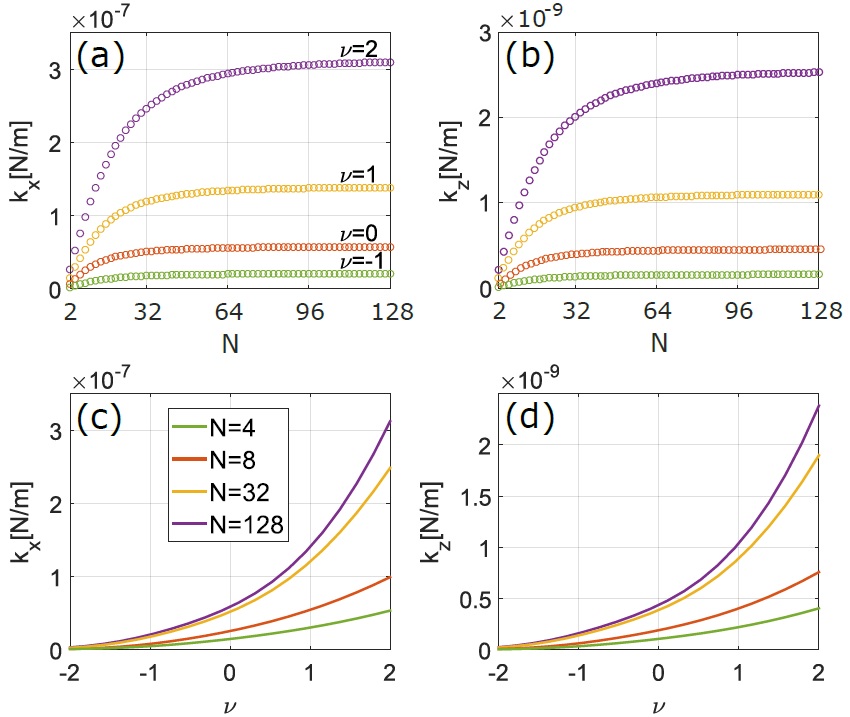}
 \caption{Trap stiffness $\kappa_{x}$ e $\kappa_{z}$ as a function of the number of beams $N$ in $(a)$ and $(b)$, and as funtion of the initial launch angle $\nu$ in  $(c)$ and $(d)$.}
 \label{Stiffness}
\end{figure}

In Fig.~\ref{Stiffness}$(a)$ and $(b)$ show the trap stiffness $\kappa_{x}$ and $\kappa_{z}$ as an function of $N$ for different initial lunch angle. We can see that both, $\kappa_{x}$ and $\kappa_{z}$, increase with the number of AiBs in the array until reaching its maximum saturation value. On the other hand, the stiffness also can be increased by increasing the launch angle as can be seen in Fig.~\ref{Stiffness}$(c)$ and $(d)$.

\emph{Trapping stability analysis} Finally, we will analyze the stability of the traps. To have a stable trap, the following conditions must be met \cite{Jiang2013}. First, the backward longitudinal gradient force must be large enough to overcome the forward scattering force, which is satisfied as we can see in Figs.~\ref{Force_distribution_different_N} and \ref{Intensity_Force}.  Second, the longitudinal gradient force must be large enough to overcome the influence of the and the gravity. In this paper, the gravity forces are much less than the maximum of longitudinal radiation force in the trap position. The third condition for stable trapping is that the potential well of the gradient force should be deep enough to overcome the kinetic energy of the Brownian particles. This condition can be expresses as follows \cite{Harada1996,Cheng2010}
\begin{equation}
R_{b}=\exp\left( U/k_{B}T\right) << 1
\label{stable_trapping}
\end{equation}
where $U$ is the potential energy of the gradient force at the trap position. $k_{B}$ is the Boltzman constant, and $T$ is the ambient temperature. In the table~\ref{Stability1} we can observed that for $\nu=0$ the particles could by stably trapped in the plane $z_{\nu_{0}}$ for $N=32$ and $N=128$, because the condition $R_{b}<<1$ is satisfied. But for the $N=4$ and $N=8$, the particle can not be stably trapped by transverse gradient force and the longitudinal gradient force, because $R_{b}$ for both is approximately equal to $1$. The table~\ref{Stability2} we show that for $N=128$ the particles could by stably trapped in the planes $z_{\nu_{2}}$, $z_{\nu_{1}}$, and $z_{\nu_{0}}$ for $\nu_{2}$, $\nu_{1}$ and $\nu_{0}$ respectively. However, the particle can not stably trapped by transverse gradient force and the longitudinal gradient force in the plane $z_{\nu_{-1}}$ for $\nu=-1$, Because $R_{b}$ for both approaches $1$.
\begin{table}[htbp]
\centering
\caption{\bf trap stability for $\nu=0$ and different values of $N$}
\begin{tabular}{ccc}
\hline
N &  $R_{b}$ for TGF &  $R_{b}$ for LGF \\
\hline
$4$ & 0.98  & 0.97  \\
$8$ & 0.97 & 0.96  \\
$32$ & $9.97\times 10^{-3}$  & $4.04\times 10^{-4}$ \\
$128$ &  $5.33\times 10^{-3}$  &  $1.45\times 10^{-4}$  \\
\hline
\end{tabular}
  \label{Stability1}
\end{table}
\begin{table}[htbp]
\centering
\caption{\bf trap stability for $N=128$ and different values of $\nu$}
\begin{tabular}{cccc}
\hline
$\nu$ & $z$ & $R_{b}$ for TGF & $R_{b}$ for LGF\\
\hline
$2$ & $z_{\nu_{2}}$ & $1.37\times 10^{-7}$ & $1.39\times 10^{-10}$ \\
$1$ & $z_{\nu_{1}}$ & $3.32\times 10^{-5}$ & $8.45\times 10^{-8}$\\
$0$ & $z_{\nu_{0}}$ & $5.33\times 10^{-3}$ & $1.45\times 10^{-4}$ \\
$-1$ & $z_{\nu_{-1}}$ & $0.26$ & $0.28$ \\
\hline
\end{tabular}
  \label{Stability2}
\end{table}

We will analyze the stability of the trap for $N=128$ and $\nu=2,1,0,-1$. For $\nu=2$ and a particle trapped at $z_{\nu_{1}}$, $R_{b}=1.37\times 10^{-7}$ for the transverse gradient force and,  $R_{b}=1.39\times 10^{-10}$ for the longitudinal gradient force. For $\nu=1$ and $\nu=0$ $R_{b}$ for the transverse gradient force is about $3.32\times 10^{-5}$ and $5.33\times 10^{-3}$, and $8.45\times 10^{-8}$ and $1.45\times 10^{-4}$ for the longitudinal gradient force respectively. \\

\h {\em\bf 4. Conclusions}  

In summary, the optical forces of AiABs exerted on Rayleigh particles composed of different Airy beams in the $N$-array and different initial launch angle ($\nu$) are investigated. The results indicate that a significant improvement in trap stability is achieved by increasing the number of beams ($N$). On the order hand, additional gain towards the stiffness is obtained by controlling the AiAB focal point through the initial launch angle ($\nu$), resulting in an increased AiAB convergence and therefore larger gradient forces, similar to using a higher NA objective in optical tweezers.  Thus, the AiABs might have great potential applications in optical trapping, optical manipulation, optical guidance and others in biological systems and nanotechnology. \\

\h {\em Acknowledgments} The authors acknowledge financial support from UFABC, CAPES, FAPESP (grant 16/19131-6) and CNPq (grant 302070/2017-6).

\end{document}